Parolini, Giuditta; Dahmen, Sílvio R.

Exploring Scientific Exchange in Agricultural Meteorology with Network Analysis

## Abstract

Network analysis is becoming increasingly relevant in the historical investigation of scientific communities and their knowledge circulation processes, because it offers the opportunity to explore and visualize connections among scientific actors on a scale qualitatively different from traditional historical methods. Temporal networks are especially suitable for this task, as they allow to investigate the evolution of scientific communities over time. In this paper we will rely on the analytical tools provided by temporal networks to examine the technical commission on agriculture (1913-1947) established by the International Meteorological Organization (IMO). By using the membership data available on this commission, we will investigate how this scientific community evolved over the decades, who were its key members, which national groups were represented, and how historical events, such as the two world wars, impacted on the work of this organization. This will give us an insight into the knowledge circulation processes of this scientific body, as the IMO was an international organization based on voluntary cooperation and its work was first and foremost the immediate consequence of the interaction among its members. In our paper we will rely on centrality measures (eigenvector, joint, and conditional centrality) to understand the structure of the commission's network, and we will constantly point out strengths and weaknesses of temporal networks in the analysis of historical data.

## 1      Introduction*

∗ **Acknowledgements**: Giuditta Parolini is grateful to the German Research Foundation (DFG) for providing funding for this research (Project No. 321660352). Sílvio R. Dahmen would like to acknowledge the financial support from the Alexander-von-Humboldt Foundation and thank

Network analysis is becoming increasingly relevant in the historical investigation of scientific communities and their knowledge circulation processes, because it offers the opportunity to explore and visualize connections among scientific actors on a scale qualitatively different from traditional historical methods.[1][2] Hundreds and even thousands of events can be brought together and examined to reconstruct the dynamics of scientific exchange. Temporal networks are especially suitable for this task, as they allow to investigate the evolution of phenomena over time.[3] The quantitative tools provided by network analysis become especially valuable, when historical sources are scarce and do not allow to reconstruct scientific exchange using traditional historian's tools and methodologies.

The case study on the history of meteorology here presented is one such example. We will examine knowledge circulation within the International Meteorological Organization (IMO), which was founded in the second half of the 19th century to promote worldwide cooperation in meteorology and climatology.[4] For several decades, before it was transformed in today's World Meteorological Organization (WMO), the IMO promoted exchange of meteorological and climatological data and knowledge across the globe via publications and regular meetings. Among its members, there were not only meteorologists and climatologists, but also scientists working in sectors heavily dependent on weather and climate knowledge, such as aviation, navigation, and agriculture.

---

Giovanna Lazzari Miotto for sharing her python expertise and helping with the code. We thank the library of the World Meteorological Organization and the Swedish Meteorological and Hydrographic Office for making available copies of the proceedings of the IMO Commission for Agricultural Meteorology. A preliminary report on this research was presented at the 13th Workshop on Historical Network Research (Akademie der Wissenschaften und der Literatur, Mainz, May 2019). We are grateful to the workshop participants for the helpful comments received there.
**Authors**: Corresponding author: Giuditta Parolini, Institut für Philosophie, Literatur-, Wissenschafts- und Technikgeschichte, Technische Universität Berlin, 10623 Berlin Germany; giuditta.parolini@tu-berlin.de;

Sílvio R. Dahmen, Instituto de Física, Universidade Federal do Rio Grande do Sul, Porto Alegre RS 91501 970 Brazil.

Primary and secondary sources available to study scientific exchange within the IMO are limited. Only a few historical accounts have been written on the IMO and they focus on its institutional developments.[5] Archival documents are also scarce, because the WMO archives only have copies of the publications produced by the IMO, but do not hold meeting minutes and correspondence that may help in reconstructing scientific interactions among the members of the organization. To make matters worse, the IMO's work was not managed only by the International Meteorological Committee, the organization's steering body, but it was also distributed across technical commissions that studied specific meteorological and climatological problems, ranging from solar radiation to weather telegraphy. Within these commissions, members collaborated to address specific issues and knowledge circulated from the national to the international level and back again towards the participating nations.

Giving the scarcity of both primary and secondary literature and the limitations of these sources, we decided to approach the study of knowledge circulation within the IMO through the quantitative tools of network analysis. As a first step towards understanding scientific exchange within the IMO, we focused on one of its technical commissions, the technical commission on agriculture, and examined how its membership evolved and developed during the first half of the twentieth century. In the paper we use membership data as a proxy to study knowledge circulation within the IMO, because information (agendas, proceedings, meeting minutes, and references to scientific publications) regularly circulated among the commission members, who, in turn, were invited to share their national best practices and scientific results, and to comment on other members' work. As the IMO was an international organization based on voluntary cooperation, its work was first and foremost the immediate consequence of the direct interaction among its members, therefore studying its membership is a key element in understanding the scientific exchange that took place.

The data used in the network analysis were extracted from publications and archival documents related to the IMO technical commission on agriculture and checked and complemented with other primary and secondary literature, as discussed in section 2. Before describing the data collection work and the resulting data set, we will briefly introduce the history and aims of the IMO technical commission on agriculture.

---

[5] On the history of the IMO, its structure and status see Ibid.; Howard Daniel, "One Hundred Years of International Co-Operation in Meteorology (1873-1973): A Historical Review," *WMO Bulletin* 22, no. 3 (1973): 156–99; Paul N. Edwards, "Meteorology as Infrastructural Globalism," *Osiris* 21, no. 1 (2006): 229–50, doi:10.1086/507143.

## 1.1    The IMO technical commission on agriculture

The IMO technical commission on agriculture was established in 1913, but its first meeting could not take place due to the outbreak of WWI. At the conclusion of the conflict, the commission was re-appointed in 1919, and in the interwar years the commission held several meetings all over Europe – in Utrecht (September 1923), Zurich (September 1926), Copenhagen (September 1929), Munich (September 1932), Danzig (August 1935), Salzburg (September 1937). The commission did not convene in Berlin in 1939, during the last IMO meeting before WWII, but it returned to work in 1946 and held its last meeting in 1947. In the early 1950s, the IMO became the WMO and its technical commissions ceased to exist. However, within the WMO, a new Commission for Agricultural Meteorology was established and still exists today.

The members of the IMO Commission for Agricultural Meteorology came from all over the world: from Europe, its colonies and protectorates, from the Americas, from Japan, and after WWII, also from China. The problems they faced were not just the collection and analysis of weather data familiar to meteorologists and climatologists. Agricultural meteorology required also to understand how weather and climate affected farming and to prepare special forecasts (of frost, hail, etc.) helpful for farmers. Therefore, the IMO technical commission on agriculture co-opted, alongside meteorologists and climatologists, also scientific experts working as geographers, agronomists, rural economists, soil scientists, plant pathologists, statisticians, etc. Agricultural meteorology, in fact, was not just a scientific pursuit, but it had an economic value for the national economies of the first half of the twentieth century. It held the promise to make farming more resilient to adverse weather events, to increase crop yields by adopting varieties better suited to local climates, and to adjust the farming calendar according to the seasonal weather, especially at key times like sowing and harvesting.

Being a member of the IMO technical commission on agriculture, therefore, offered the opportunity to discuss national practices in agricultural meteorology, learn about foreign experiences in the field, and agree on common strategies for developing this new research area that had both a scientific and an economic value. Thus, by investigating how the commission's membership changed, who were the key members of this body, which national groups were represented and how this changed over time, network analysis gives us the opportunity to understand how knowledge circulation developed during the first half of the twentieth century within this IMO commission.

In principle, the analysis here described can be replicated for all the main IMO technical commissions, which remained in activity for several decades and regularly made available member lists. In turn, a better understanding of these commissions can pave the way for a more general network analysis of the IMO.

## 2      Membership of the IMO technical commission on agriculture: data sources and general overview

The main data sources used in this paper are the proceedings of the meetings held by the IMO (proceedings of the meetings held by the International Meteorological Committee, proceedings of the conferences of the directors of meteorological services, proceedings of the meetings held by the IMO technical commission on agriculture). These proceedings are available through libraries and we were able to collect the volumes relevant to the study of the IMO technical commission on agriculture. Appendix H in Hendrick Cannegieter's institutional history of the IMO was also helpful to reconstruct the history of the technical commission on agriculture.[6] Archival documents related to the work of the IMO are not conserved by the WMO archives, but it was possible to gather some archival sources relevant to understanding the work of the IMO and its technical commission on agriculture through the scientific papers of scientists, such as the meteorologist William Napier Shaw, who was president of the International Meteorological Committee and a member of the IMO technical commission on agriculture, and the statistician R. A. Fisher, who was a member of the technical commission on agriculture during the 1920s. These archival documents were mainly helpful to reconstruct how the work of the technical commission on agriculture was organized and the role that president and secretary of the commission had in managing the knowledge circulation among the commission's members.

### 2.1    Data sources and information extracted for the network analysis

We reconstructed the membership of the IMO technical commission on agriculture using the member lists published in the IMO general proceedings and in the proceedings of the commission. All the data sources are listed in Table 1.[7] The member lists provide basic information, such as name, affiliation (not always), and role of the member in the commission. The role in the commission – member, secretary/vice-president, president – is a key element in our network analysis of the commission, because the president, with the help of the secretary, was at the center of the correspondence network on which the work of the commission was based. During the first half of the twentieth century only a few members had the opportunity to attend the meetings in person, but all could participate in the work of the commission by correspondence. All the members received a copy of the agenda, minutes, and proceedings distributed by president and secretary, and they could contribute to the work of the commission by mailing in return letters with their comments and proposals, and copies of their scientific papers.

---

[6] Cannegieter, "The History of the International Meteorological Organization 1872-1951," 198–203.

The member lists given in the IMO publications were examined for consistency and correctness. Year of birth and death for the members were always recorded, whenever available in biographical sources. In this way it was possible to realize, for instance, that the Portuguese agronomist Filipe Eduardo de Almeida Figueiredo, who died in 1930, was wrongly listed as a member of the commission both in 1932 and 1935. In compiling the data for the commission's membership, inconsistencies in the spelling of names were corrected, and affiliations checked against biographical information provided by reference works and obituaries. For the 1923 meeting in Utrecht, there is not a complete list of the commission's members, but the proceedings cite all the new members added to the commission in that year and some of the previous members. The list for 1923 was compiled, therefore, according to a *continuity principle*. To the names explicitly mentioned in the proceedings, we added also the names of the members who were listed both in the previous (1921) and in the following (1926) meeting of the commission.

In building the network we took into account both the role of the individual members (president, secretary, etc.) and the nation (historically conceived) where they worked, because these two factors are critical to understanding how scientific exchange developed within this international body. The role is not just a formal label, but it identifies privileged routes of information transfer within the commission. The president had the task to prepare a report of the activity carried out by the commission since the previous meeting and to circulate it among the members. He was also responsible for inviting members to meetings, prepare the meeting agenda, circulate the list of new members and the minutes after each meeting, and organize the publication of the proceedings. Members corresponded with the president to contribute to the work of the commission. In the early 1920s, the role of secretary was introduced to support the work done by the president and, with only a few exceptions, all the presidents of the commission served as secretaries before moving on to the role of president.[8] From the historical point of view, there is no doubt that the relationship between president and secretary was a privileged one and that president and secretary were in contact with every member of the commission.

---

[7] The full dataset of the commission's membership is available for consultation here: Parolini, Giuditta (2020), "Membership of the IMO Commission for Agricultural Meteorology (1913-1947)", Mendeley Data, v1, http://dx.doi.org/10.17632/pds6tz443t.1.

The other piece of information we used in our network analysis is the nation where the members worked. Members who worked in the same nation had a privileged relationship, because they belonged to the same intellectual circles and could meet and cooperate also on a national level. We used an historical, rather than a geographical understanding of nation. That is, we counted as French also the members working in French colonies, such as French Indochina, because they had constant interactions with the scientific circles in their home country. Similarly, members working in British, Dutch or Belgian colonies were considered part of the scientific circles in their homeland. However, we considered members working in British protectorates, such as Canada or South Africa, that had more scientific autonomy than colonial officers, as part of independent national groups.

| Year | Meeting | Reference | Information provided |
|------|---------|-----------|----------------------|
| 1913 | Meeting of the International Meteorological Committee (Rome) | *Bericht über die Versammlung des Internationalen Meteorologischen Komitees (Rom, 1913)*, Berlin 1913, Behrend & Co | Member name Member role |
| 1919 | Conference of Directors of the Meteorological Services (Paris) | *Report of the Conference of Directors of the Meteorological Services (Paris, 1919)*, London 1920, Meteorological Office | List with: Member name Affiliation Member role |
| 1921 | Meeting of the International Meteorological | *Report of the 11th Meeting of the International Meteorological Committee* | List with: Member name Affiliation Member role |

[8] In the proceedings of the last commission's meeting in 1947, this support role is labelled as vice-president rather than secretary.

| | | | |
|---|---|---|---|
| | Committee (London) | *(London, 1921)*, London 1922, HMSO | |
| 1923 | Conference of Directors of the Meteorological Services and meeting of the International Meteorological Committee (Utrecht) | *Report of the International Meteorological Conference of Directors and of the Meeting of the International Meteorological Committee (Utrecht, 1923)*, Utrecht 1924, Koninklijk Nederlandsch Meteorologisch Instituut, No. 112, Kemink & Zoon | No list available, but members of the commission are mentioned in the report on the 1st meeting of the IMO Commission for Agricultural Meteorology |
| 1926 | Meeting of the IMO Commission for Agricultural Meteorology | *Commission de Météorologie Agricole: Procès verbaux de la 2ième réunion (Zürich, 1926)*, Statens Meteorologisk-Hydrografiska Anstalt 256, Stockholm, 1927 | List with: Member name Affiliation Member role |
| 1929 | Meeting of the IMO Commission for Agricultural Meteorology (Copenhagen) | *Commission de Météorologie Agricole: Procès verbaux de la 3ème réunion (Copenhague, 1929)*, Statens Meteorologisk-Hydrografiska Anstalt 276, Stockholm, 1929 | List with: Member name Affiliation Member role |
| 1932 | Meeting of the IMO Commission for Agricultural Meteorology (Munich) | *Commission de Météorologie Agricole: Procès verbaux des seances de Munich (Munich, 1932)*, Organisation Météorologique Internationale, Utrecht, 1933 | List with: Member name Affiliation Member role |
| 1935 | Meeting of the IMO Commission for Agricultural | *Kommission für Landwirtschaftliche Meteorologie: Protokolle der Tagung in Danzig*, Secretariat de | List with: Member name Affiliation Member role |

| | | |
|---|---|---|
| | Meteorology (Danzig) | l'Organisation Météorologique International No. 24, Imprimerie Edouard Ijdo, Leyde, 1936 | |
| 1937 | Meeting of the IMO Commission for Agricultural Meteorology (Salzburg) | *Kommission für Landwirtschaftliche Meteorologie: Protokolle der Tagung in Salzburg*, Secretariat de l'Organisation Météorologique International No. 36, Imprimerie Edouard Ijdo, Leyde, 1938 | List with: Member name Affiliation Member role |
| 1946 | Extraordinary conference of Directors of the Meteorological Services and meeting of the International Meteorological Committee (London) | *Conférence extraordinaire des directeurs (Londres, 25 fevrier -2 mars 1946), Comité Météorologique International (Londres, 2 mars 1946)*, Organisation Météorologique International No. 52, Imprimerie La Concorde, Lausanne, 1946 | List with: Member name Location (city) Member role |
| 1947 | Meeting of the IMO Commission for Agricultural Meteorology (Toronto) | *Commission for Agricultural Meteorology: Abridged Final Report (Toronto, 1947)*, Organisation Météorologique International No. 63, Imprimerie La Concorde, Lausanne, 1949 | List with: Member name Location (city and nation) Member role |

**Table 1.** Data sources.[9]

---

[9] The IMO technical commission on agriculture did not meet in 1939, when the International Meteorological Committee held its last meeting in Berlin before the outbreak of WWII. The proceedings of the meeting include, however, a short report written by the commission president, the Dutch Cornelis Braak. The report provides information on the members who left the commission and the new ones, whose admission was planned during the meeting of

## 2.2 The members of the IMO commission: a brief overview in numbers

From 1913 to 1947, 132 people became members of the IMO technical commission on agriculture. This overall number, however, helps little to understand how the membership of the commission changed over time, because the membership figures fluctuated significantly. Before WWI, the commission only had a handful of members, less than ten. But it began to grow steadily during the interwar years and reached the highest number of members, 68, in 1935 (Figure 1). After WWII, the number of members dropped to less than forty, but the data available for 1947, the last year on which information is available, suggest that the commission membership quickly began to increase again.

Among the technical commissions nominated by the IMO, the Commission on Agricultural Meteorology was one of the largest, as it co-opted not only meteorologists and climatologists, but also scientific experts working in fields such as agriculture and botany. The constant growth of its membership during the first half of the twentieth century is also not surprising, as agricultural meteorology was becoming more and more popular within scientific and institutional circles connected, in particular, to farming and agricultural research, and articles and textbooks on the subject multiplied during this period.

---

the commission in 1941. Such meeting never took place due to WWII. By comparing the information in Braak's report with the member list for 1937, it becomes evident that the proceedings of the 1937 meeting, published in 1938, already took into account large part of the changes communicated in the 1939 report regarding the members who left the commission. The only exceptions are the Finnish meteorologist Gustaf Melander, who died in 1938, and the British meteorologist G. C. Simpson, who retired from his position at the Meteorological Office in September 1938.

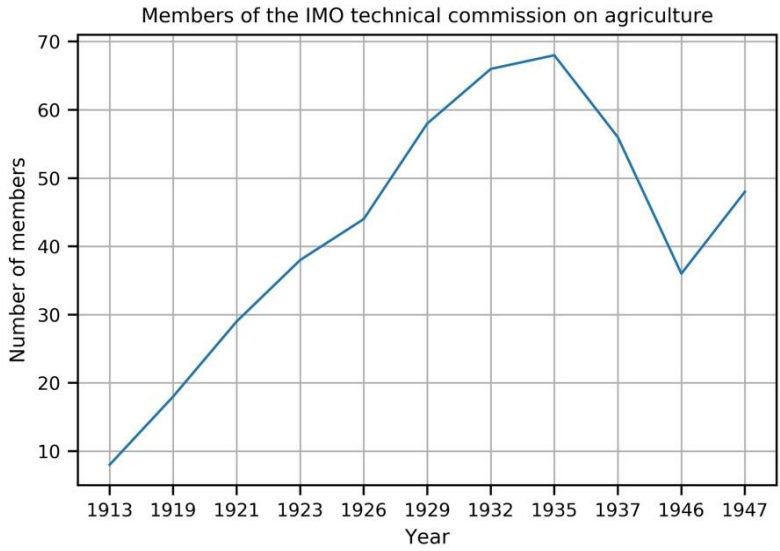

**Figure 1**. Variation over time of the membership figures for the IMO Commission on Agricultural Meteorology.

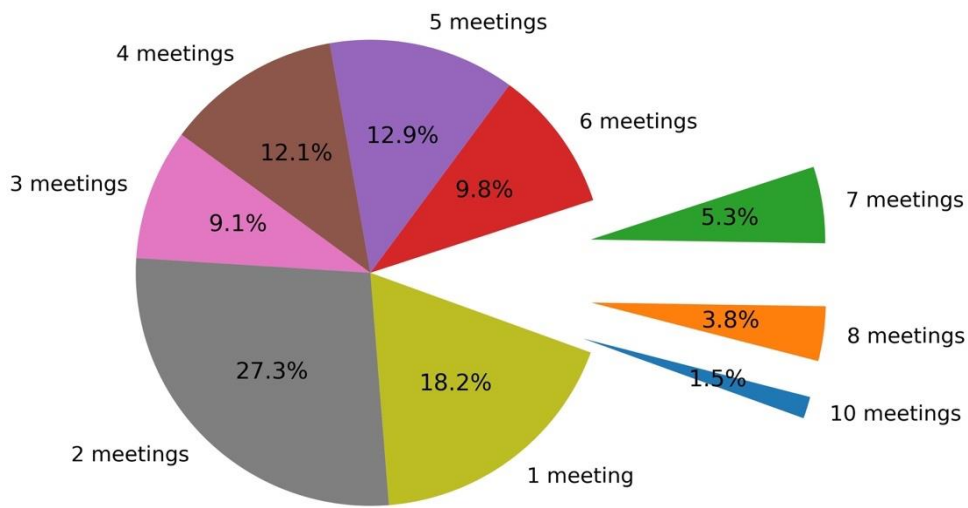

**Figure 2**. Members' participation in the IMO Commission for Agricultural Meteorology.

Membership in the commission was a long-term commitment for the majority of the people who took part in the activity of this scientific body (Figure 2). Less than twenty per cent of the commission members are mentioned in just one member list, while almost fifty per cent of the overall members are mentioned in two to four lists. Considering the periodicity of the meetings organized by the IMO, the members of its technical commission on agriculture remained part of it on average for three to six years. This is certainly an interesting piece of information considering the scattered geographic distribution of the members and the complex historical events, not to mention the two world wars, that the IMO had to deal with, and that significantly hindered scientific exchange.

Long-term members, i.e. the people listed in seven to ten member lists, are about ten per cent of the total number of members. Among them we find European members, such as the Italian agronomist Girolamo Azzi and the Norwegian meteorologist Hans Theodor Hesselberg, but also scientists based out of Europe, such as the agronomist Paul Carton working in French Indochina, the meteorologist Abraham James Connor working in Canada, and the scientist Takematsu Okada working in Japan. This suggests that participation in the commission was not necessarily hindered by geographic distance and that the correspondence network on which the commission work was based worked satisfactorily also for its extra-European members.[10] If geographic distance was not necessarily an obstacle, belonging to a certain nation could be an hindrance to become a member of the commission. After WWI, for instance, German and Austrian scientists were excluded from taking part in the work of international scientific institutions such as the IMO, due to the support they had given to the war policies of their countries and they re-joined the ranks of the technical commission on agriculture only in 1926. Similarly, Russian scientists only returned to the commission after the revolution and the creation of the Soviet Union in the early 1920s.

Nations and nationalities are therefore a crucial element to take into account while examining the membership of the IMO commission. In studying it, we are facilitated by scientists' restricted mobility in the first half of the twentieth century. With the exception of a few cases – the Australian geographer Thomas Griffith Taylor, who moved from his home country to the United States and Canada, and the Swiss meteorologist Jean Lugeon, who worked also in Poland – the members of the IMO technical commission on agriculture continued to work in the same nation, often also in the same institution, throughout the years.

---

[10] On the extra-European networks of knowledge exchange established by the IMO Commission for Agricultural Meteorology see Giuditta Parolini, "Building Networks of Knowledge Exchange in Agricultural Meteorology: The Agro-Meteorological Service in French Indochina," *History of Meteorology*, forthcoming.

Figure 3 reproduces the distribution of the commission's members according to their nation, while Table 2 provides a global overview for each nation and clarifies how national groups have been formed. As evident from the heatmap, on average each nation had one or two representatives in the commission. Significant exceptions to this trend are given only by France, Great Britain, and Germany. In the interwar years, French scientists, based in the homeland and the colonies, were the biggest national group within the commission. In 1932, 14 out of the 66 members of the commission (over twenty per cent), were French. Agricultural meteorology indeed became a popular discipline in France earlier than in other countries – the first president of the commission, the meteorologist Charles Alfred Angot, was French –, but perhaps, to counterbalance this striking numerical presence, the commission only elected one French president and one French secretary during its history.

By examining the heatmap, it becomes evident that only three nations, Canada, Italy, and France, had representatives in the commission throughout its history. Many more, such as the Netherlands, Great Britain, Brazil, Norway, Spain, and Japan, were often, but not always, represented within this scientific body. These fluctuations can certainly be attributed to the availability of national experts interested in agricultural meteorology, but also to the changing political equilibria of the first half of the twentieth century. After WWII, for instance, several nations, such as Turkey, Egypt, Palestine, China, and Peru, which had not been previously represented, entered in the commission and continued their membership within the WMO technical commission on agriculture. In the following section, we will introduce our network analysis of the commission membership and use it to investigate in detail the trends we found by inspecting the data.

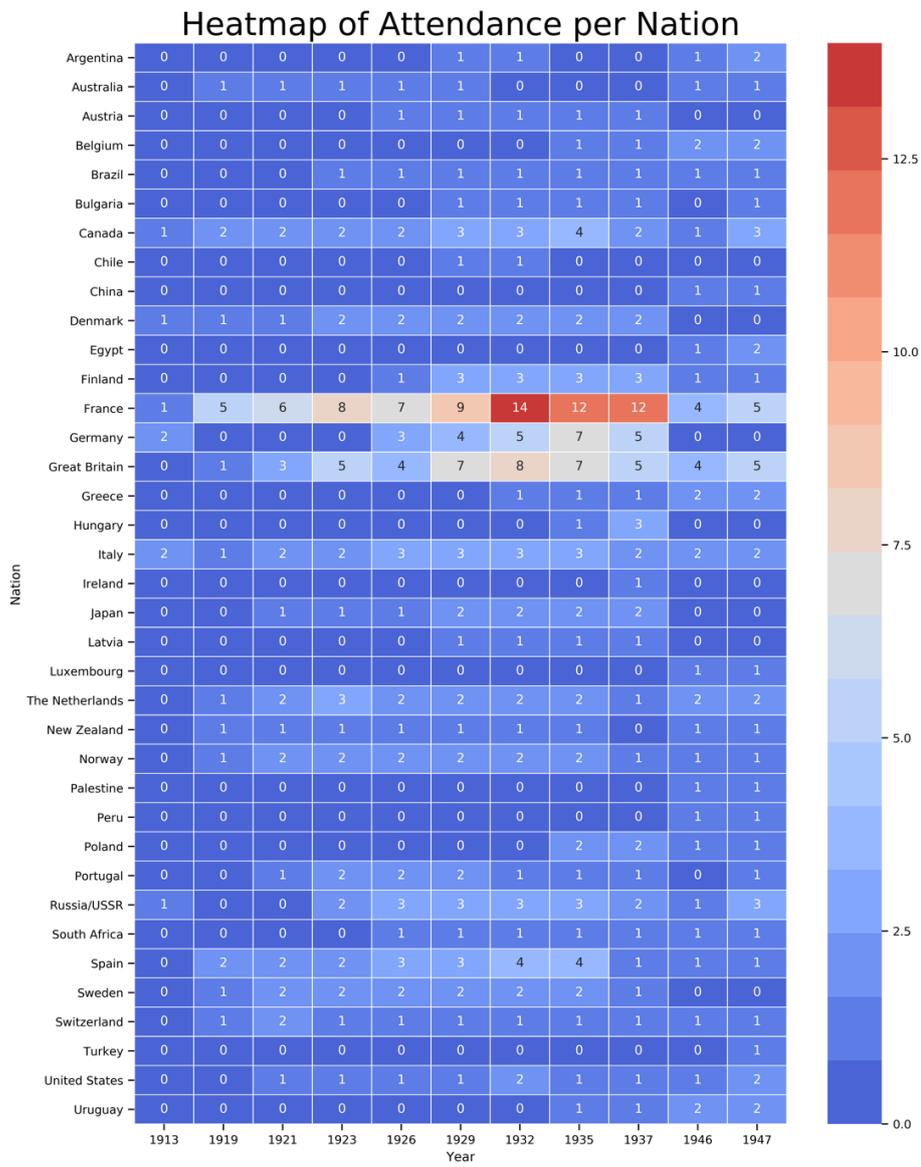

**Figure 3.** Members of the IMO Commission for Agricultural Meteorology distributed according to nation.

|     | Nation                                                        | Members' number (total) |
| --- | ------------------------------------------------------------- | --- |
| 1   | Argentina                                                     | 3   |
| 2   | Australia                                                     | 2   |
| 3   | Austria                                                       | 2   |
| 4   | Belgium (including Belgian Congo)                             | 3   |
| 5   | Brazil                                                        | 2   |
| 6   | Bulgaria                                                      | 1   |
| 7   | Canada                                                        | 8   |
| 8   | Chile                                                         | 1   |
| 9   | China                                                         | 1   |
| 10  | Denmark                                                       | 3   |
| 11  | Egypt                                                         | 2   |
| 12  | Finland                                                       | 3   |
| 13  | France (including French Syria, Indochina, and North Africa)  | 18  |
| 14  | Germany                                                       | 10  |
| 15  | Great Britain (including India, Ceylon, British East Africa)  | 13  |
| 16  | Greece                                                        | 2   |
| 17  | Hungary                                                       | 3   |
| 18  | Ireland                                                       | 1   |
| 19  | Italy                                                         | 5   |
| 20  | Japan                                                         | 2   |
| 21  | Latvia                                                        | 1   |
| 22  | Luxembourg                                                    | 1   |
| 23  | The Netherlands (including Dutch East Indies)                 | 7   |
| 24  | New Zealand                                                   | 2   |
| 25  | Norway                                                        | 3   |
| 26  | Palestine                                                     | 1   |
| 27  | Peru                                                          | 1   |
| 28  | Poland                                                        | 2   |
| 29  | Portugal                                                      | 4   |
| 30  | Russia/USSR                                                   | 7   |
| 31  | South Africa                                                  | 2   |
| 32  | Spain                                                         | 5   |
| 33  | Sweden                                                        | 3   |
| 34  | Switzerland                                                   | 3   |
| 35  | Turkey                                                        | 1   |
| 36  | United States                                                 | 5   |
| 37  | Uruguay                                                       | 2   |

**Table 2.** Nations represented in the IMO technical commission on agriculture.

# 3  Network analysis of the IMO technical commission on agriculture

## 3.1  Building the network

Information on the activity of the IMO commission circulated among all the people who were listed as members in a specific year, but president and secretary of the commission had special roles in managing and redistributing this information among the members. We decided to capture this dynamic of scientific exchange by building a multilayer temporal network (Figure 4). Each layer corresponds to one of the years for which data on members are available (1913, 1919, 1921, etc.).

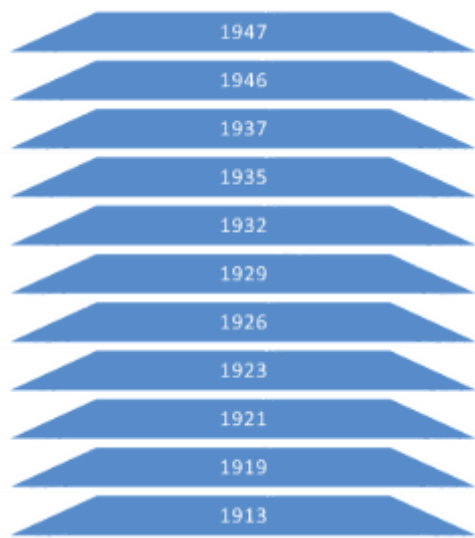

**Figure 4**. Temporal layers on which the network of the IMO Commission for Agricultural Meteorology is built.

The nodes of the network are the individual commission members and an edge is drawn between people who are recorded as members in the same year. Within each layer, therefore, every member is connected to every other member (i.e. we have a fully connected graph). This implies that the underlying topology of the network, which is key to understanding the relevance of nodes, cannot be extracted from a simple visual analysis, even though the participants in each meeting were always less than 70, because all nodes of a fully connected network are topologically equivalent. It is through the assignment of *weights* to the edges that the equivalence is broken and the standard centrality measures (see below) can be used to investigate the network of the IMO commission.

The weights (Table 3) range from 4 to 0.5 and have been assigned considering both the role of each member within the commission and their nationality, because scientific exchange heavily depended on these two factors. The highest weight, 4.0, is attributed to the connection between president and secretary, because their collaboration was central to the information exchange within the commission. All the other members (for a certain year) are connected to the president with weight 3.0 and to the secretary with weight 2.0, because president and secretary were gatekeepers (with a different level of importance) in the information exchange. If president and member, or secretary and member, belonged to the same nation, we account for this by adding an additional 0.5 to the weight (3.5 instead of 3.0 and 2.5 instead of 2.0, respectively), because these people had greater opportunities of information exchange between themselves. A lower weight, 1.0, is attributed to the edges that connect members working in the same nation and that were neither president nor secretary, because the role in the commission had a greater relevance than the national affiliation. The lowest weight, 0.5, is given to the people who did not work in the same nation, but were members at the same time. This connection accounts for the scientific exchange that took place among members who potentially never met in person nor exchanged letters, but nonetheless had access to the scientific information exchanged within the commission through proceedings and minutes of the meetings. The IMO network is not directed, as we are able only to investigate if members A and B had an opportunity to exchange information, but not who was the sender and who the receiver of information. To clarify weight assignation, we report, as an example, the data for 1913, the meeting with the fewest members.

| Weight | Applies to |
|--------|-----------|
| 4.0 | President-Secretary |
| 3.5 | President-Member (same nation) |
| 3.0 | President-Member (different nation) |
| 2.5 | Secretary-Member (same nation) |
| 2.0 | Secretary-Member (different nation) |
| 1.0 | Member-Member (same nation) |
| 0.5 | Member-Member (different nation) |

**Table 3.** Network Weights.

| Surname | First name | Nation | Role |
|---------|-----------|--------|------|
| Angot | (Charles) Alfred | France | President |
| Börnstein | Richard Leopold | Germany-Prussia | Member |
| Brounov | Peter Ivanovich | Russia | Member |
| Dop | Louis | Italy | Member |
| Hergesell | Hugo | Germany-Bavaria | Member |
| Palazzo | Luigi | Italy | Member |
| Ryder | Carl H. | Denmark | Member |
| Stupart | (Robert) Frédérick | Canada | Member |

**Table 4.** IMO Commission members in 1913.

| Member A | Member B | Weight | Edge type |
|----------|----------|--------|-----------|
| Angot (France) | Börnstein (Germany) | 3 | President-Member (different nation) |
| Angot (France) | Brounov (Russia) | 3 | President-Member (different nation) |
| Angot (France) | Dop (Italy) | 3 | President-Member (different nation) |
| Angot (France) | Hergesell (Germany) | 3 | President-Member (different nation) |
| Angot (France) | Palazzo (Italy) | 3 | President-Member (different nation) |
| Angot (France) | Ryder (Denmark) | 3 | President-Member (different nation) |
| Angot (France) | Stupart (Canada) | 3 | President-Member (different nation) |
| Dop (Italy) | Palazzo (Italy) | 1 | Member-Member (same nation) |
| Börnstein (Germany) | Hergesell (Germany) | 1 | Member-Member (same nation) |
| Brounov (Russia) | Börnstein (Germany) | 0.5 | Member-Member (different nation) |
| Dop (Italy) | Börnstein (Germany) | 0.5 | Member-Member (different nation) |
| Palazzo (Italy) | Börnstein (Germany) | 0.5 | Member-Member (different nation) |
| Ryder (Denmark) | Börnstein (Germany) | 0.5 | Member-Member (different nation) |
| Stupart (Canada) | Börnstein (Germany) | 0.5 | Member-Member (different nation) |
| Dop (Italy) | Brounov (Russia) | 0.5 | Member-Member (different nation) |

| Hergesell (Germany) | Brounov (Russia) | 0.5 | Member-Member (different nation) |
|---|---|---|---|
| Palazzo (Italy) | Brounov (Russia) | 0.5 | Member-Member (different nation) |
| Ryder (Denmark) | Brounov (Russia) | 0.5 | Member-Member (different nation) |
| Stupart (Canada) | Brounov (Russia) | 0.5 | Member-Member (different nation) |
| Hergesell (Germany) | Dop (Italy) | 0.5 | Member-Member (different nation) |
| Ryder (Denmark) | Dop (Italy) | 0.5 | Member-Member (different nation) |
| Stupart (Canada) | Dop (Italy) | 0.5 | Member-Member (different nation) |
| Palazzo (Italy) | Hergesell (Germany) | 0.5 | Member-Member (different nation) |
| Ryder (Denmark) | Hergesell (Germany) | 0.5 | Member-Member (different nation) |
| Stupart (Canada) | Hergesell (Germany) | 0.5 | Member-Member (different nation) |
| Ryder (Denmark) | Palazzo (Italy) | 0.5 | Member-Member (different nation) |
| Stupart (Canada) | Palazzo (Italy) | 0.5 | Member-Member (different nation) |
| Stupart (Canada) | Ryder (Denmark) | 0.5 | Member-Member (different nation) |

**Table 5.** Data for the network analysis (Year = 1913). [11]

## 3.2   Centrality measures used to evaluate the temporal network

From a theoretical point of view, the standard measures for evaluating the importance of a node in a network are based on the concept of *centrality*.[12] As the name implies, centrality measures how relevant (i.e. central) a node is. This relevance can be evaluated according to different criteria, therefore there are several types of centrality measures. One example is the so-called *betweenness centrality*, that is the importance of a node in connecting different parts of a network which are not directly connected themselves. The most straightforward centrality measure and the easiest to perform computationally is the *degree centrality*, which is nothing but the number of connections a node has. However, the importance of a node is not so much measured by the number of nodes it is connected too, but by the quality of these nodes. For this reason, a very common measure used to ascertain the relevance of a network node is its *eigenvector centrality*, that is, a measure which takes into account how well-connected a node is to other well-connected nodes. From all possible centrality measures one can compute in analyzing a network, the most appropriate one is contingent on the research question being asked.

---

[11] For each year, the number of possible combinations is given by the formula [$n(n$-1)/2], where $n$ is the number of members for that year. For 1913 the possible combinations are [8(8-1)/2] = 28.

[12] Michele Benzi and Christine Klymko, "A Matrix Analysis of Different Centrality Measures," *SIAM Journal on Matrix Analysis and Applications* 36, no. 2 (2013): 686–706; Linton C. Freeman, "Centrality in Social Networks: Conceptual Clarification," *Social Networks* 1, no. 3 (1979): 215–39.

Centrality measures depend strongly on the dynamics of the network, since nodes may come and go, shifting the underlying topology and consequently those measures which depend on it. Due to the computational complexity of time-dependent networks as well as open issues regarding appropriate methods to treat them, most literature on the subject leaves temporal aspects aside and treat networks as static. In many cases where a temporal sequence of networks representing different realizations of the same network is known, one may opt to use an aggregate network, that is, summing all different networks as if all connections depicted took place within the same time frame.[13] In this paper we opted instead for a full-blown temporal version of the network under study as this depicts in a more realistic way the relevance of the members over the decades in which the IMO technical commission on agriculture was in existence. We will be using three main centrality measures, namely the *eigenvector centrality*, the *joint centrality*, and the *conditional centrality*.[14]

The **eigenvector centrality**, also called prestige score, is a measure of the importance of a node in the structure of a network. In order to have a high value of eigenvector centrality it does not suffice for a node to have many connections, but it has to be connected to nodes that are also well-connected. This means that it is not the number of connections that count but the quality of the nodes one is attached to. The ubiquitous PageRank algorithm used for finding popular websites is an example of an eigenvector-centrality-based algorithm. The name 'eigenvector centrality' stems from the fact that it is computed using the

---

eigenvector associated to the highest eigenvalue of the adjacency matrix used to represent the network mathematically.[15]

The **joint centrality** is a measure of the importance of a node at a given time step weighted by its values along the whole story of the network. The joint centrality is the equivalent of the eigenvector centrality in a temporal setting, that is when different realizations of the network interact with each other. Its value may thus differ significantly from the eigenvector centrality, because the latter represents an image of the network at a given time without considering other time steps, while the joint centrality captures the relevance of a node throughout the 'temporal history' of the network. This is possible by connecting different realizations of a network via a coupling constant, which preserves the identity of its nodes but carries the net of influences in one network to the next one. However, one should bear in mind that it is not the absolute values of centralities that count, since as the network changes, so the number of nodes and connections may change. For this reason, it is important also to have a measure of the centrality of a node at a given time step in relation to the other nodes at that time step. This is given by 'normalizing' the joint centrality by the sum of all joint centralities at a given time step, thus obtaining the so-called **conditional centrality**.

We will be using the measures just mentioned to examine quantitatively the relationships existing among the nodes of the network of the IMO technical commission on agriculture. Before doing so, however, we want to consider the visualization of the network we have built, as this visual inspection is relevant in understanding whether the model we built is able to reproduce key events in the history of the commission.

## 3.3    Network visualization

The visualization of the temporal network of the IMO Commission for Agricultural Meteorology is displayed below in Figure 5. In this visualization, the commission members, who are the nodes of the network, are represented by dots of different colors, while the edges are represented as gray arrows. The size of each dot is proportional to the number of edges that depart/arrive on that dot and the weight of those edges. Please note that this is a temporal visualization, therefore the edges visible are drawn taking into account the member interactions across all the time layers used to build the network. White is the standard color we used for nodes, but we colored differently the nodes

---

representing members who belonged to the nations with the greatest participation within the history of the commission. The graph legend explains the color code adopted. The choice to emphasize these members through a color code helps to identify at a glance how national participation changed and shifted across the history of the commission.

By examining the network visualization, we can identify key historical events that affected the work of this scientific body. At a glance, one can identify three main areas within the graph. The members who contributed to the foundation of the commission in 1913 (see Table 4), but did not take part in its activity after WWI, are displayed at the bottom of Figure 5. The largest group at the center of the image is formed by the members who took part in the work of the commission during the interwar years. Within this group, it is possible to identify a smaller subgroup of members who took part in the activity of the commission also after WWII and are a go-between the pre- and post-WWII history of the commission. The last group displayed at the top of the image is formed by the members who joined the commission only after WWII. Within each group, the position of the members and the length of the edges is automatically assigned by the software to produce a readable graph, where each node and its label are visible. It goes without saying that the graph here presented is not the only possible representation of the IMO commission network. Equivalent visualizations could also be drawn, for instance with the post-WWII group at the bottom and the pre-WWI group at the top. These representations would have the same topological structure and the centrality measures would not change in any way.

The overall structure of the network here obtained is in agreement with what we know about the history of the IMO. The two world wars caused sharp discontinuities in the work of this organization, because all the meetings traditionally held by the IMO and its technical commissions were cancelled and for years its members could not regularly exchange scientific information. In the aftermath of the conflict the organization's work had to be rebuilt and the membership renewed, because many people had left their office or died during the war years.

In the case of the IMO technical commission on agriculture, this is especially evident with reference to WWII. During the interwar years the commission membership constantly grew, and more and more nations joined this scientific body and regularly participated in its activity. Among these nations, France can be singled out for the remarkable number of members that joined the commission and remained part of it for several years. From the network visualization, it then become evident the contrast to the post-WWII situation. Not only were the members fewer than during the interwar years, but they were also in large part newly recruited scientists and, interestingly, among them just one was affiliated to a French institution. Only fifteen members transitioned from pre- to post-WWII and among them we can find a few of the scientists, such as the Italian agronomist Girolamo Azzi and the Norwegian Theodore Hesselberg, who spent more time than anyone else in the commission.

However, a qualitative analysis of the network visualization does not allow to study in detail the relevance of the long-term members within the commission and its scientific exchanges. The same is true for the president and secretaries of the commission. We know that they had a key role in the information exchange of the organization, but we cannot rank them, nor study them in detail, just by looking at the network visualization. The graph portrays the evolution of the commission over more than three decades, but presidents and secretaries remained in charge for a few years only, therefore their individual relevance is blended in the global story told by the graph. We can examine the importance of specific people and national groups within the commission only by resorting to the centrality measures mentioned above. These measures will be the analytical tools we will use in the following section to discuss how specific people and nations shaped scientific exchange in the IMO Commission for Agricultural Meteorology.

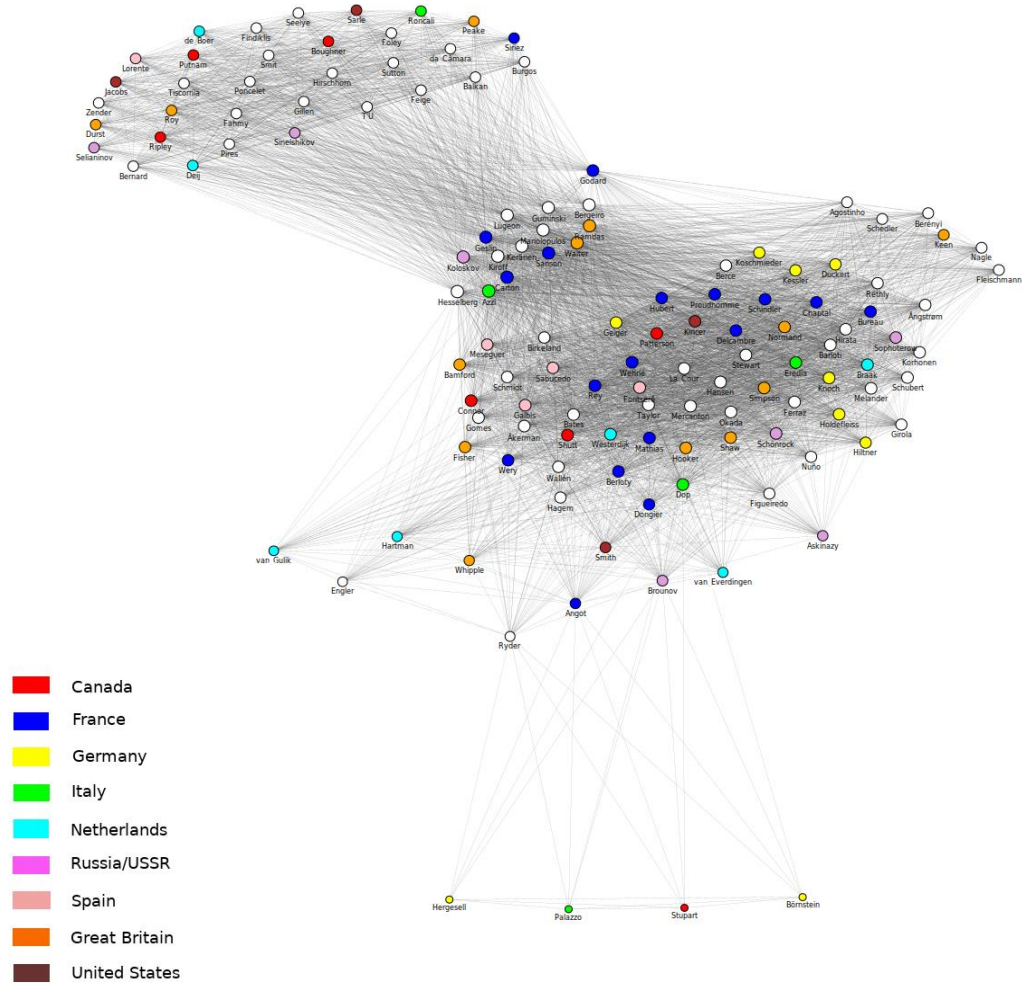

**Figure 5**. Temporal network of the IMO Commission for Agricultural Meteorology (1913-1947).

# 4   People, nations, and scientific exchange

## 4.1   Presidents and secretaries/vice-presidents

The key people in managing scientific exchange within the IMO technical commission on agriculture were its presidents and secretaries/vice-presidents, because they were in correspondence with all the commission members, they used the suggestions received from members to prepare the agenda for the meetings, and after each conference they curated the publication of the proceedings and distributed them. As they constantly supported the knowledge exchange within the scientific body, we have made presidents and secretaries/vice-presidents crucial nodes in our network of the IMO commission, and we now want to examine their individual relevance by using the centrality measures described above.

To facilitate the discussion, we will be considering plots of the different centrality measures. In the plots each time layer is represented on the horizontal axis by a number ranging from 1 to 11. These numbers correspond to each year for which membership information is available (1—1913, ..., 11—1947). We will examine how the centrality values change over time and what this can tell us about the role of presidents and secretaries/vice-presidents in the exchange of scientific information within the commission. To correctly interpret these graphs, please bear in mind two relevant aspects of temporal methods in network theory:[16]

- The network matrices must always have positive eigenvalues, since centralities have to be positive by definition. This can only be achieved if matrices are symmetrical. In a temporal network, a symmetrical matrix implies that there is **backward causation**, i.e. what is happening in year 1935, for instance, influences what happened in 1913. This is unavoidable when centrality measures, which were originally developed for static networks, are extended to temporal networks, but it must be taken into account while interpreting the results.

- **When a node disappears from the network**, in our case, for instance, when a president leaves his office, **the centrality values of that specific node decline sharply, but do not become equal to zero in forthcoming years**, because there is a coupling between the temporal layers that form our network. Intuitively, we can explain this property pointing out that even if a node disappears, at least a few of the nodes he was connected to remain in the network and carry forward the influence of the lost node.

---

[16] Appendix A provides more information on the mathematical theory of temporal networks and the computation of eigenvector, joint, and conditional centralities.

## Eigenvector centrality

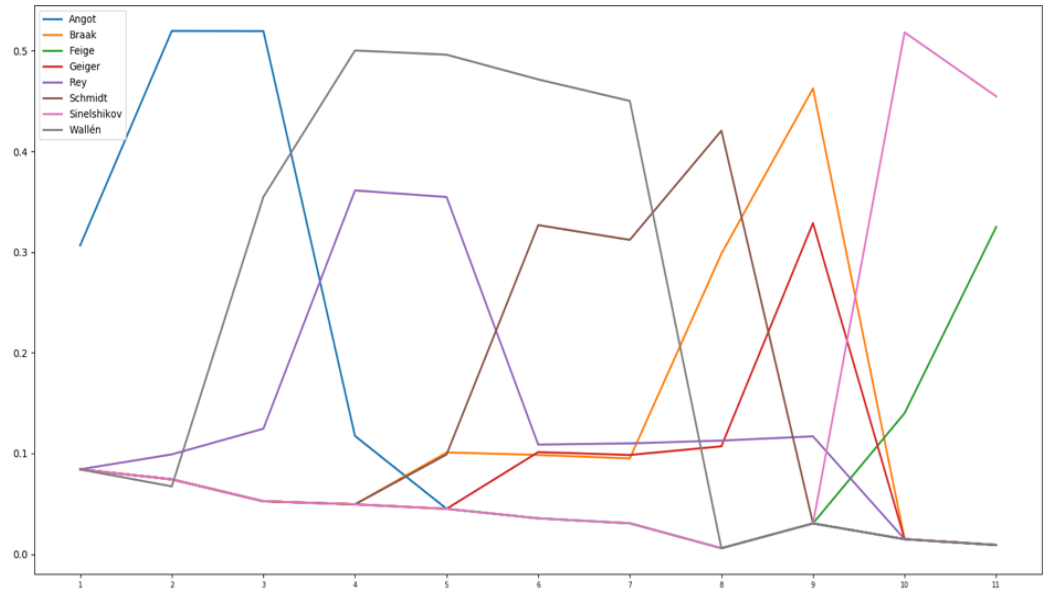

**Figure 6.** Plot of the eigenvector centrality for presidents and secretaries/vice-presidents.

The eigenvector centrality (see Sect. 3.2) is the standard measure for evaluating the relevance of a node in a network. In our case, the eigenvector centrality offers a view of the relevance of presidents and secretaries/vice-presidents, as if they were examined year by year. This becomes clear considering, for instance, how the eigenvector centrality of the first president of the commission, the French meteorologist Charles Alfred Angot, changes over time. Angot was one of the founding members of the commission and managed it since 1913. After his retirement from the directorship of the French Meteorological Bureau in the early 1920s, he left the role of commission president, but not the commission itself where he remained as a member until his death in 1924. We see therefore that while Angot's eigenvector centrality increases from year 1 to year 2 hand in hand with the increase in the number of the commission members and remains stable from year 2 to year 3, then it has a sharp decline in correspondence of year 4, when he becomes only a member, and from then onwards it constantly decreases, although it never becomes zero for the reasons mentioned above.

At a glance, it is evident that the Swedish meteorologist and hydrologist Axel Wilhelm Wallén was central to the functioning of the commission for a much longer time than any other president or secretary. His centrality starts increasing in 1921, when he is elected secretary, reaches a maximum in 1923, when he becomes president, and then it remains stable until 1932. The sharp decline in Wallén's centrality value in 1935 is due to his sudden death. At the death of Wallén, the commission elected two more presidents before WWII, the Austrian meteorologist Wilhelm Mathäus Schmidt and the Dutch climatologist Cornelis Braak. Both held office for a short time only. The former because he died in 1936 and the second because his presidency was interrupted by WWII and, by the time the IMO re-started his activities after the conflict, he was nearing retirement. Like Wallén, both Schmidt and Braak had previously acted as secretaries of the commission and their plots increase in two steps: the first raise corresponds to the change in role from member to secretary and the second to the switch from secretary to president.

In the interwar years, the IMO technical commission on agriculture had two more secretaries we did not mention so far, the French agronomist Pierre Rey and the German meteorologist Rudolf Geiger. Both never achieved the role of president and their plots, as expected, always remain below the presidents' own eigenvalue centrality plots. In the case of Rey, we can see his move from ordinary member to secretary in 1923, the same year in which Axel Wallén's left the position of secretary to take up office as president. Rey remained secretary until 1926, when he left this role to Schmidt, but continued to be a member of the commission until WWII has evident by the shape of his plot. Similarly, Geiger was a member of the commission for a few years, before he took the role of secretary just before WWII and did not rejoin the commission after the war.

In the aftermath of the conflict, many new members joined the commission. The president, the Soviet hydrometeorologist Victor Vasilyevich Sinelshikov, and vice-president, the Jewish meteorologist Rudolf Feige, who managed the commission in the last few years of its existence, had never been members before, as evidenced by their centrality plots that significantly raise only for the years 1946 and 1947. These plots are also a good example of the 'backward causation effect' in temporal networks. Both Sinelshikov and Feige, in fact, have a non-null eigenvector centrality even in the years (1913 to 1937) during which they were not members of the commission, but, as mentioned above, nodes they were connected too in the late 1940s (see Figure 5, members group connecting before and after WWII), had been part of the history of the commission also in the 1920s and 1930s.

## Joint centrality

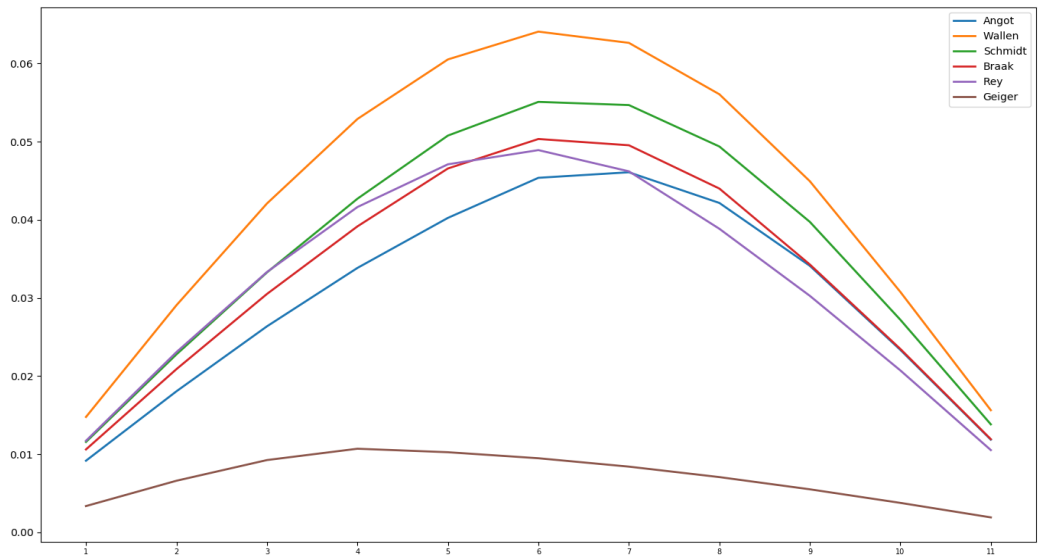

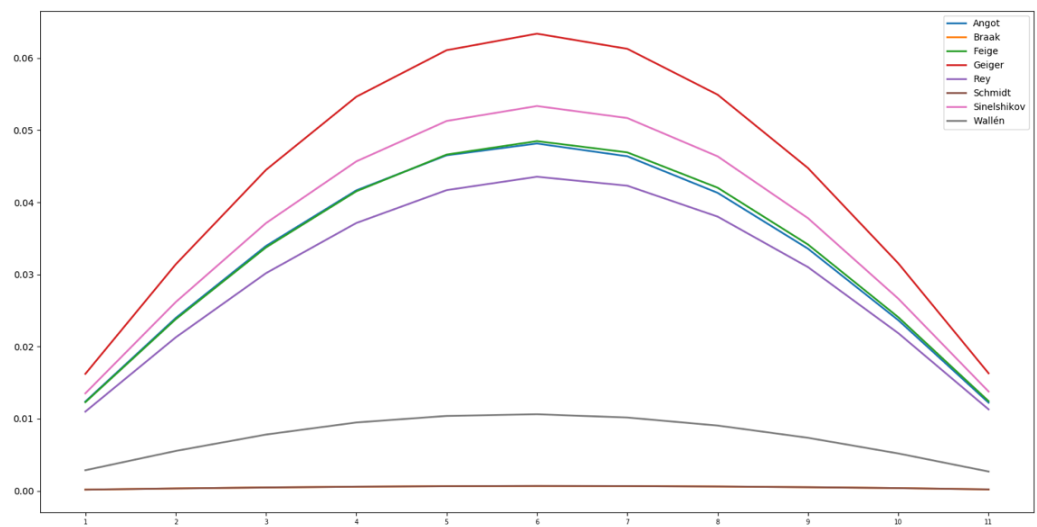

**Figure 7a,7b**. Plot of the joint centrality for presidents and secretaries/vice-presidents (a: 1913-1937, b: 1913-1947).

The joint centrality (see Section 3.2) gives the opportunity to evaluate the relative importance of presidents and secretaries during the entire history of the commission, because it takes into account the coupling existing between different time layers. The two images above are joint centrality plots with a strong temporal coupling drawn respectively for presidents and secretaries between 1913 to 1937 (Figure 7a) and for all presidents and secretaries/vice-presidents (Figure 7b). In Figure 7a, Wallén's relevance in shaping the work of this scientific body during the interwar years is confirmed. Not only did Wallén remain in office longer than any other president, but he also took part in the expansion phase of the commission and the number of members almost doubled during his time. Schmidt and Braak, who held the office of president in the 1930s, also have higher values of conditional centralities compared, for instance, to Angot, the first president, whose membership in the commission already ended in the early 1920s. The lowest joint centrality value is Geiger's, who was the last secretary before WWII.

The joint centrality plots for presidents and secretaries have a similar shape, almost an inverted parabola. This happens because the computation of this centrality measure is influenced by the number of members in the network and has therefore a peak in the years (late 1920s-early 1930s), in which the membership of the commission was larger. As proved by Taylor and colleagues, in the limit of extremely strong temporal coupling, the joint centrality tends to the shape of a sine wave and our plots are just a portion of the theoretical curve.[17]

It is quite interesting to see how the joint centrality plots radically change by including in the graph also Feige and Sinelshikov (Figure 7b). Paradoxically, in this case, the highest centrality is attained by Geiger, followed by Sinelshikov and Feige. Wallén's joint centrality becomes very low, and the same can be said about the joint centrality of Schmidt, who had the second highest value in Figure 7a. In order to explain these counterintuitive results, we need to return to the network visualization (Figure 5). As we already pointed out, WWI and WWII generated strong discontinuities in the scientific work done by the commission and, before and after, we have quite significant changes in the membership of the commission. In our case, while WWI did not have a relevant effect, because the members of the commission before the conflict were very few, WWII produced a radical change. This change helps us to explain the results obtained in Figure 7b and why the joint centrality is unable to describe properly our network.

Mathematically, the joint centrality is obtained from the combination of 11 adjacency matrices, one for each time layer. These matrices are built into one large matrix where each separate block is connected to the other. In this global matrix each member of the commission is identified with himself/herself in each realization of the network and each realization of the network is coupled to the other by a factor. The joint centrality is then the 'eigenvector centrality' of this combined matrix and measures the importance of a commission member based on the importance of the members he/she is connected to. Due to the symmetry requirements in the combined matrix, Feige and Sinelshikov are mathematically represented since the first year of the commission and benefit from the 'connections of their connections'. That is, the group of members who remained in the commission before and after WWII (see Figure 5) are also connecting Feige and Sinelshikov to the previous members of this scientific body. Among these 'connecting members' many were long-term members of the commission, therefore well-connected to the majority of the members, and thus they significantly raise the joint centrality value of the post-WWII newcomers. In our case the only conclusion we can draw is that with strong discontinuities in time, caution must be applied in using the joint centrality and the best approach is to produce plots that refer only to homogeneous time-phases, which, in our case, are mainly pre- and post-WWII.

---

[17] Taylor et al., "Eigenvector-Based Centrality Measures for Temporal Networks".

## Conditional centrality

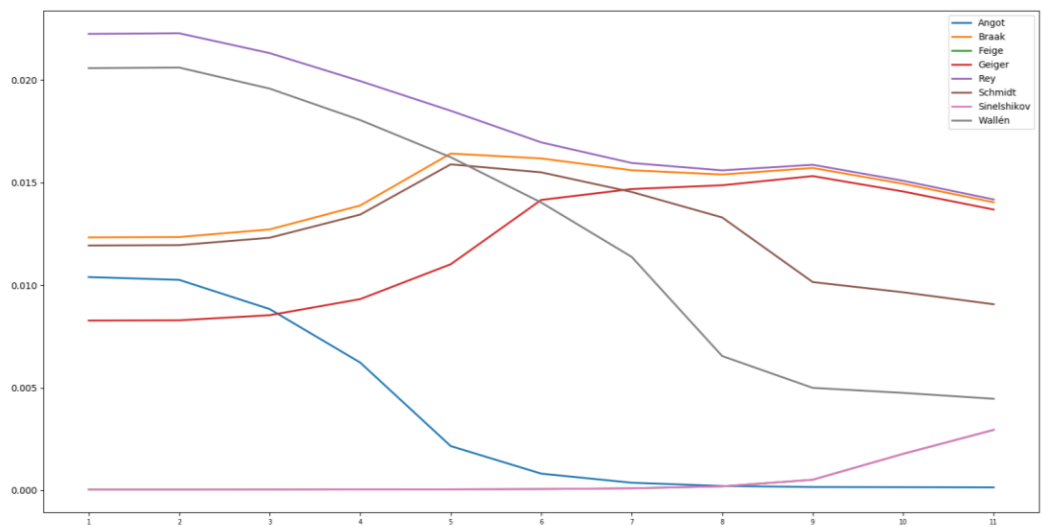

**Figure 8**. Plot of the conditional centrality for presidents and secretaries/vice-presidents.

The conditional centrality allows us to focus on each time layer, because it is obtained by normalizing the joint centrality by the centralities of all the members in a given year. Due to this normalization, the measure overcomes the issues that we discussed above in using the joint centrality. Interestingly the highest conditional centrality score is not attained by a president, but by a secretary of the commission, the French agronomist Pierre Rey, who became a member of the commission in 1919 and such remained until 1937. Rey held the office of secretary from 1923 until 1928. During these years, the commission constantly expanded its membership increasing the number of scientists with whom Rey had an opportunity to exchange information. In addition, before WWII, the group of French scientists taking part in the commission's work became bigger and bigger. This further contributes to raise Rey's centrality due to the increased weight we attributed to edges connecting scientists working in the same nation.

This example helps to understand the value of network analysis compared to traditional historical methodologies. Rey was an agronomist only known in French agricultural circles for his contributions to agricultural meteorology.[18] In comparison to the commission presidents and even to some of its long-term members, who were famous meteorologists and climatologists very active in international circles, he was not a prominent scientist. Yet, as our network analysis points out, in the bigger picture of the commission's scientific exchange he had an important role, because his constant attendance brought him in contact with all the members who took part in the commission's work in the interwar years.

A mention should also be made of Geiger, the last secretary elected before WWII. While his eigenvector centrality was smaller, even at its peak, than the eigenvector centrality of any other president or secretary, his conditional centrality is quite high, because his membership of the commission coincided with the years during which more members were present. This is a point that is important to emphasize in order to interpret this centrality measure. Aside from the weights that we attributed to the edges, a key factor in determining the centrality of a node is the number of connections to well-connected nodes. It follows that the members who participated in the 1930s, when the commission had more than sixty members on average, have a higher chance to be connected to many well-connected nodes. Therefore, we have an higher conditional centrality for Geiger, compared, for instance, to Angot's conditional centrality, because the latter only participated in early meetings, in which the number of members remained under forty, and because Angot was connected also to people (the pre-WWI members) who did not reappear after the conflict or quickly disappeared in the early 1920s.

| Year | President | Secretary/Vice-president |
|------|-----------|--------------------------|

| 1913 | (Charles) Alfred **Angot** | -------------- |
|------|------------------------------|----------------|
| 1919 | (Charles) Alfred **Angot** | -------------- |
| 1921 | (Charles) Alfred **Angot** | Axel Wilhelm **Wallén** |
| 1923 | Axel Wilhelm **Wallén** | Pierre **Rey** |
| 1926 | Axel Wilhelm **Wallén** | Pierre **Rey** |
| 1929 | Axel Wilhelm **Wallén** | Wilhelm Mathäus **Schmidt** |
| 1932 | Axel Wilhelm **Wallén** | Wilhelm Mathäus **Schmidt** |
| 1935 | Wilhelm Mathäus **Schmidt** | Cornelis **Braak** |
| 1937 | Cornelis **Braak** | Rudolf **Geiger** |
| 1946 | Victor Vasilyevich **Sinelshikov** | -------------- |
| 1947 | Victor Vasilyevich **Sinelshikov** | Rudolf **Feige** |

**Table 6.** Presidents and secretaries/vice-presidents of the IMO commission.

## 4.2 Long-term members

Alongside presidents and secretaries of the commission, we decided to investigate also the role of the long-term members, that is the scientists who participated in the work of the technical commission for agriculture for long time (Table 7).[19] Among this group of thirteen people, only one had for a few years the role of secretary, Pierre Rey, and, as we have seen above, his centrality values across the whole history of the commission are quite significant, even though he never became president.[20] We expect that also for the other long-term members the centrality measures will point us to interesting results in our temporal network, and we will be reviewing these measures one by one, as we did before for the presidents and secretaries/vice-presidents of the commission.

---

[19] We selected the members who took part in seven or more meetings. They correspond to the exploded slices in Figure 2 and represent about ten per cent of the global membership of the commission.

[20] As we already examined Rey in the previous section, we did not consider him in the analysis of the long-term members.

## Eigenvector centrality

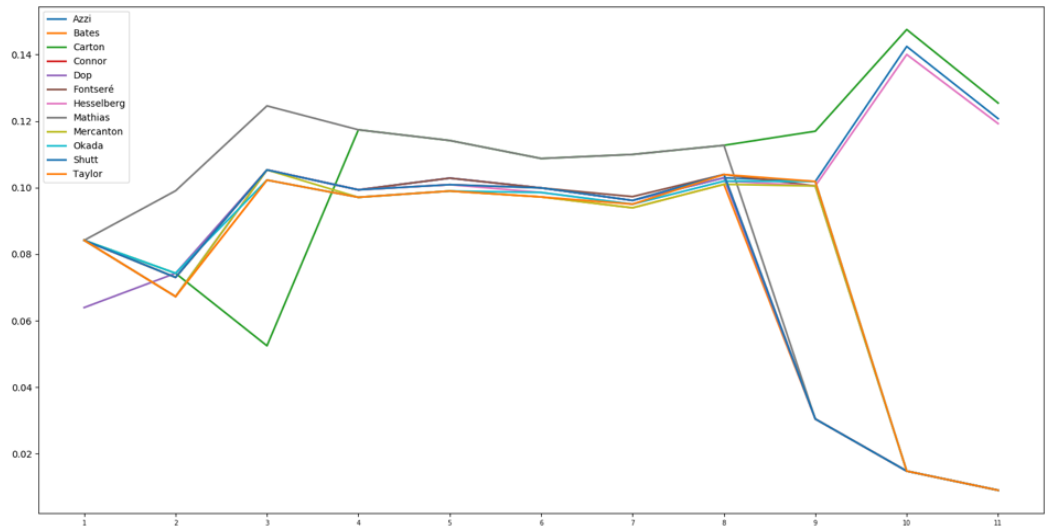

**Figure 9**. Plot of the eigenvalue centrality for long-term members.

The eigenvector centrality of the long-term members is significantly lower (about four times lower) than the eigenvector centrality of presidents and secretaries/vice-presidents. This result is not surprising as the weight system we have chosen for our temporal network emphasizes the role of presidents and secretaries/vice-presidents in the commission. All the long-term members have similar values of eigenvector centrality and these values remain rather constant over time. The same will be true also for joint and conditional centrality values. However, the members who belonged to the strongest national group, the French one, such as the agronomist Paul Carton and the meteorologist Émile Mathias, reach higher values. The weight difference we used for people working in the same nation is evidently sufficient to achieve the result we had in mind, that is to make nationality a key point in our network alongside the role in the commission.

The two members who took part in the work of the commission for longer time were the Italian agronomist Girolamo Azzi and the Norwegian meteorologist Theodor Hesselberg. Their eigenvector centrality values, however, become significantly different from the ones of the majority of the long-term members only after WWII. We interpret this result as the outcome of two factors. As remarked before, WWII created a strong discontinuity in the history of the commission and Azzi, Hesselberg, and Carton are the only members of the pre-WWII period that still are active in the commission in the 1940s. Hence, their dominant position. However, one must also keep into account that while Carton was a member of a very large national group before the war, after the conflict, the group of French scientists participating in the commission was much reduced, and also scientists who worked in countries with a limited number of participants, such as Italy and Norway, could gain higher centrality values.

From the historical point of view, it is interesting to remark that several of the long-term members of the commission were not based in Europe, even though all the meetings of the commission except one took place there.[21] Paul Carton worked in French Indochina (today's Vietnam), the meteorologist Daniel Bates in New Zealand, the geographer Thomas Taylor moved from Australia to the United States and eventually to Canada, the meteorologist Takematsu Okada was based in Japan, and the agricultural chemist Frank Shutt in Canada. If these scientists remained members of the commission during such a long time, evidently the correspondence network on which the work of the commission was based functioned well. Several of them never had an opportunity to participate in the commission meetings and yet, their remarks and suggestions can be traced in the proceedings of the meetings.[22]

---

[21] The last meeting of the commission was held in Toronto.

[22] One example is Paul Carton, who was one of the most active members of the commission. See, for instance, Parolini, "Building Networks of Knowledge Exchange in Agricultural Meteorology: The Agro-Meteorological Service in French Indochina".

## Joint centrality

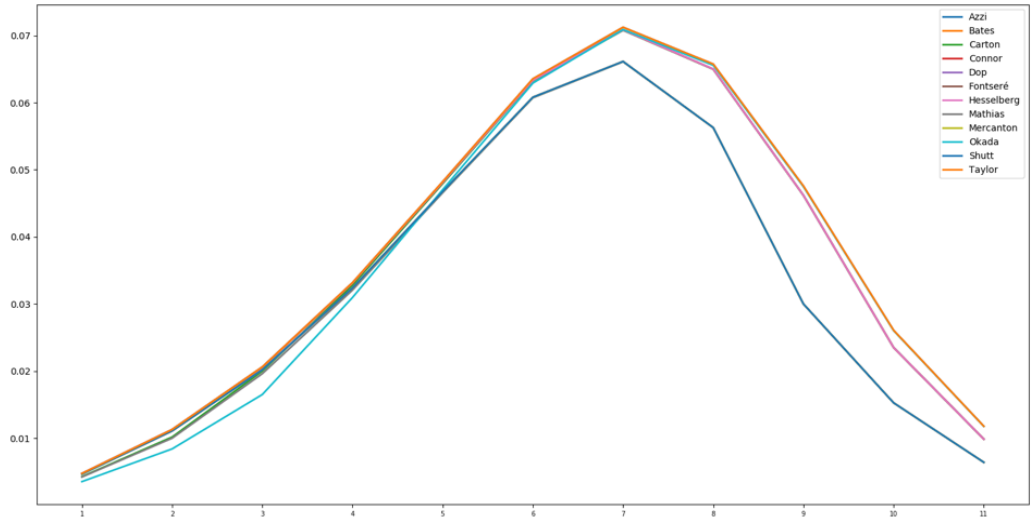

**Figure 10**. Plot of the joint centrality for long-term members.

As mentioned above, in our discussion of the joint centrality for presidents and secretaries/vice-presidents, this measure offers the opportunity to evaluate the relative importance of members during the whole history of the commission. When we compare the joint centrality values of the long-term members with those of presidents and secretaries/vice-presidents, we find very similar values. This suggests that our assumption, i.e. that these long-term members had a special importance in the history of the commission, is correct and that their permanence over the years makes them key nodes in our network.

Even for long-term members the joint centrality plots have a similar shape. The maximum value of each plot can be found in correspondence of year 1932, while the corresponding peak for the joint centrality plots of presidents and secretaries/vice-presidents was in year 1929. It is not easy to interpret the reason of this shift. Possibly the later peak of the centrality graph has to do with the fact that all the long-term members participated in the work of the commission during the interwar years, at least, while many presidents and secretaries only remained in a charge for a few years. Again, the discontinuity created by WWII in our temporal network may also contribute to this result, as before and after WWII the number of long-term members sharply decreases from thirteen to three.

## Conditional centrality

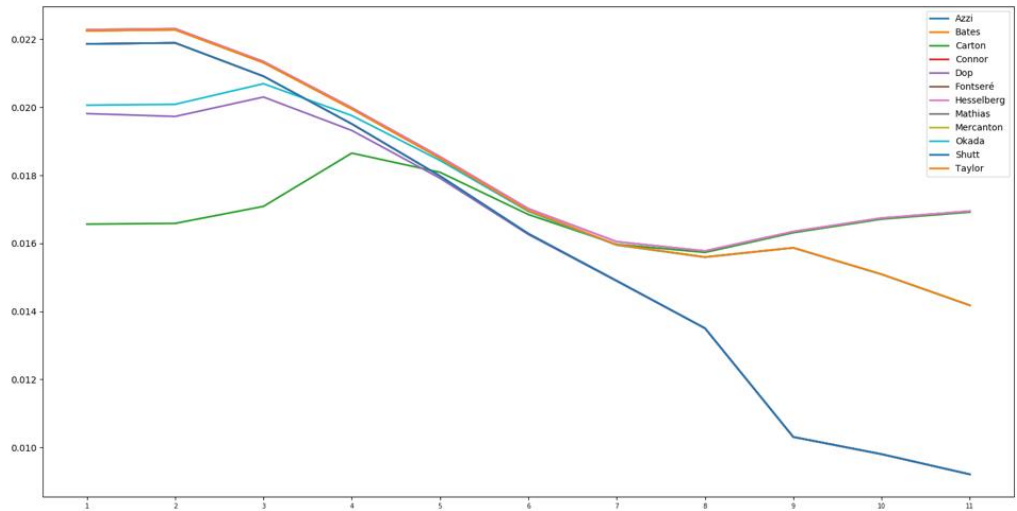

**Figure 11**: Plot of the conditional centrality for long-term members.

Similarly to the joint centrality, also for the conditional centrality the values of long-term members are comparable to the values reached by presidents and secretaries/vice-presidents, and the plots' shape has analogies to the plots obtained above for presidents and secretaries/vice-presidents. As the plots are remarkably close and often overlap, it is difficult to follow the conditional centrality of individual members. Yet, a few features still emerge, and we would like to briefly comment on that. We have already discussed the issue of back causation, intrinsic to the analysis of temporal networks, and here we can see a clear example of that. The only long-term member who took part in the work of the commission during Year 1 (1913) is the French economist Louis Dop, based in Rome at the International Institute of Agriculture. Yet, other long-term members, e.g. the Japanese Okada and the Australian geographer Taylor, have even higher values of conditional centrality in 1913. Evidently, we have to interpret this result with caution and see it as a consequence of the conditional centrality values that these members gained later in the history of the commission.

It may also be surprising to realize that the conditional centrality has a constant decrease during the interwar years, but this feature can be explained taking into account the constant expansion of the commission membership during the 1920s and 1930s: due to the higher number of members, the conditional centrality of each diminishes because this is a normalized measure. A closer look at the plot, suggests immediately that the negative trend becomes especially noticeable in 1935-1937, the years with the highest number of members. There is, however, for a few members a new phase of increase in the centrality values after WWII. This happens for the three members – Carton, Azzi, and Hesselberg – who remained in the commission after WWII, when almost all the other members were new to the work of the commission.

| Years | Name | Profession and nation |
|---|---|---|
| 1919-1947 | Girolamo **Azzi** | Agronomist, Italy |
| 1919-1947 | (Hans) Theodor **Hesselberg** | Meteorologist, Norway |
| 1923-1947 | Paul **Carton** | Agronomist, France |
| 1919-1937 | Eduardo **Fontseré i Riba** | Meteorologist, Spain |
| 1919-1937 | Paul Louis **Mercanton** | Meteorologist, Switzerland |
| 1919-1937 | Pierre **Rey** | Agronomist, France |
| 1919-1937 | Thomas Griffith **Taylor** | Geographer, Australia/US/Canada |
| 1919-1935 | Daniel Cross **Bates** | Meteorologist, New Zealand |
| 1913; 1921-1935 | Louis **Dop** | Rural economist, Italy |
| 1919-1935 | José **Galbis** | Geographer, Spain |
| 1919-1935 | Émile **Mathias** | Meteorologist, France |
| 1921-1937 | Takematsu **Okada** | Meteorologist, Japan |
| 1919-1935 | Frank Thomas **Shutt** | Agricultural chemist, Canada |

**Table 7.** Long-term members of the IMO technical commission on agriculture.

## 4.3    National groups in an age of conflict

In the first half of the twentieth century, two world wars, revolutions, and diplomatic crises constantly transformed alliances between nations and hindered international scientific collaboration.[23] In the work of the IMO and its technical commission on agriculture, the world wars marked strong discontinuities in membership and altered national equilibria, for instance, due to the ban of German scientists after WWI. It is important, therefore, to use the centrality measures computed in our network analysis also to understand how national participation in the Commission for Agricultural Meteorology changed during the history of this scientific body.

---

[23] Giuditta Parolini, "Rebuilding International Cooperation in Meteorology after World War I: The Case of Agricultural Meteorology," *Acta Historica Leopoldina*, forthcoming.

In order to evaluate the importance of national groups in the network analysis, we summed all the centrality values obtained for the individual members who belonged to the same nation and considered the resulting value as the centrality value for that nation. We will discuss now the eigenvector, joint, and conditional centralities obtained according to this procedure.

### Eigenvector centrality

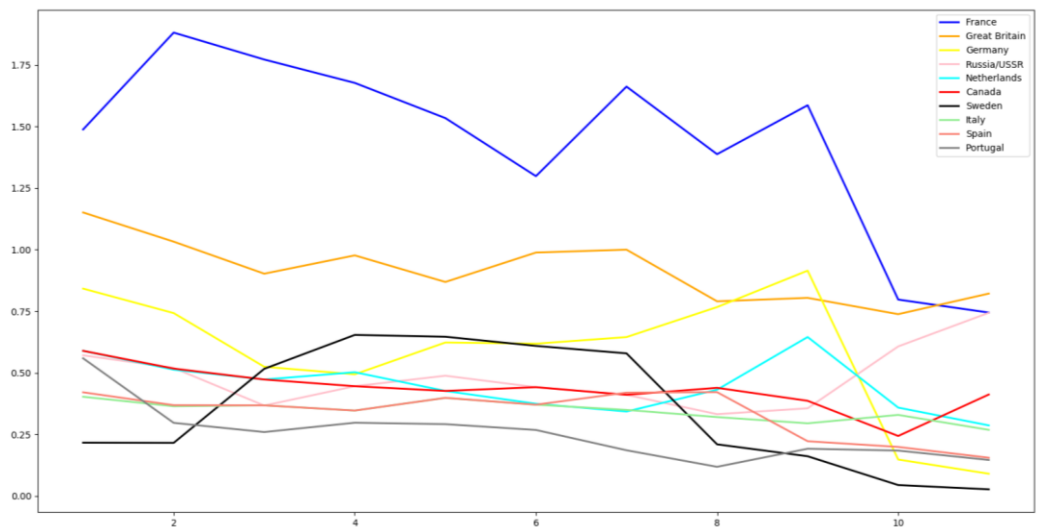

**Figure 12.** Plot of the eigenvector centrality for nations.

France had the largest members' group in the IMO Commission for Agricultural Meteorology (Figure 3). In total, 18 scientists working in France and its colonies joined the commission (i.e. 14% of the total membership). They were both meteorologists and agronomists, and worked in observatories, meteorological services, agronomy institutes, and in the colonial academies and services. The large number of French members was justified by the established tradition France had in agricultural meteorology. This tradition did not concern only the homeland, but also the colonies, because the French colonial administration considered science and technology as key factors in the promotion of colonial economies.[24] Not surprisingly, one of the most active members of the IMO Commission for Agricultural Meteorology was the French agronomist Paul Carton, based in Indochina and responsible for the creation and management of the agrometeorological service in this French colony.[25]

France is the nation with the highest eigenvector centrality values, even though only one French president (Angot) and one French secretary (Rey) were elected during the entire history of the commission. Our choice to compute nations' centrality scores as sums of members' centralities is therefore effective in revealing the real importance of national groups, and it is not biased by our weight system that attributes more relevance to presidents and secretaries/vice-presidents. This conclusion is confirmed by the eigenvector centrality values for Great Britain, the nation that follows (at a distance) France. British members never held an administrative role in the commission, but they were numerous (10% of the global membership) and constantly present throughout the history of the commission. Again, the existence of a large colonial empire with institutions interested in agricultural meteorology justifies the large numbers reached by the British group.[26]

Germany is the third country in order of relevance that can be singled out in the eigenvector centrality plot. With ten members during the entire history of the commission, i.e. 8% of the global membership, Germany's national relevance is clearly visible in the graph. However, compared for instance to the case of Great Britain, it is evident that the presence of German scientists within the commission fluctuated significantly over the years. We already mentioned the ban of German scientists after WWI. By looking at the plot, it is evident that similarly, after WWII, the German members drastically diminished, and a quick check of the membership data reveals that there were no German members in either 1946 or 1947. The German national group is different from those of France and Great Britain discussed above for an additional reason. Germany lost its colonial empire after WWI and the large participation in the commission's activities was not caused by the contemporary attendance of scientists working in the homeland

---

and in the colonies. All the German members worked for agricultural, forestry, and meteorological institutions based in Germany, and often conducted research in the microclimatology of plant growth that was a main interest in German agricultural meteorology.

The eigenvector centrality plots of the other seven nations displayed in Figure 12 are concentrated in the lower part of the graph. We will not discuss them individually, as their values are all very close, but we would like to point out a few interesting features. The first one is the growth of the eigenvector centrality values for Sweden during the presidency of Axel Wallén. As discussed above, Wallén was the most influential president of the commission in the interwar years and the prestige of Sweden, which had only three members in the entire history of the commission (i.e. 2% of the global membership), is positively affected by the key role acquired by one of its national representatives. A similar behavior is evident for the Netherlands (seven members in total, i.e. 5% of the global membership) whose eigenvector centrality has a peak in correspondence of Cornelis Braak's presidency in 1937. Even in the case of Germany, the peak is reached in correspondence of year 1937, when the only German secretary of the commission, Rudolph Geiger, was in office. This suggests that while for the largest groups, France and Great Britain, with participation percentages superior or equal to 10%, the number of members is the main factor, for smaller groups also the role of specific members contributes significantly to raise the eigenvector centrality values.

**Joint centrality**

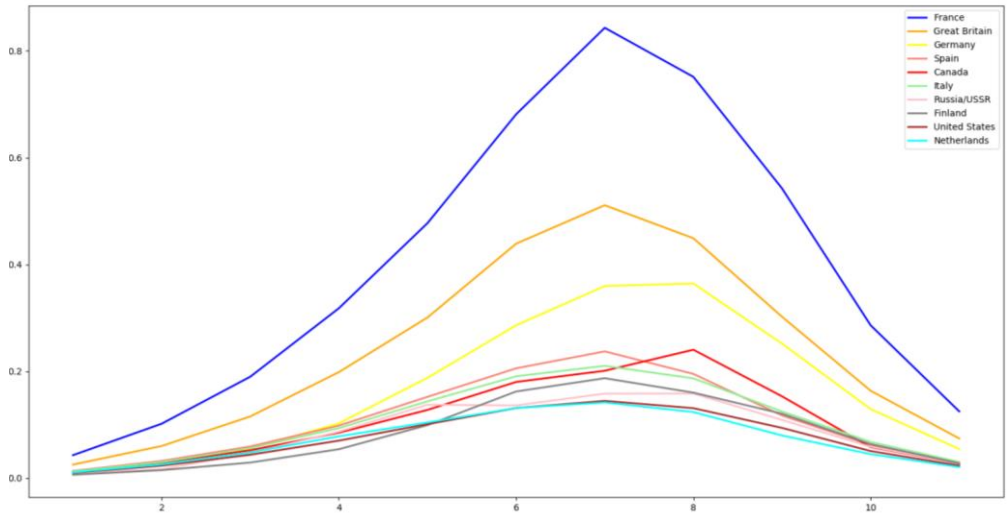

**Figure 13.** Plot of the joint centrality for nations.

The joint centrality plots are in agreement with the information provided by the eigenvector centrality values for nations. Again, the nations with the highest number of members, France, Great Britain, and Germany, have values recognizably higher than all the other nations displayed in the plot. The plots follow the sine-shaped curve that was already discussed with reference to the joint centrality values for presidents and secretaries/vice-presidents, and for long-term members.

It is interesting to point out that the other nations displayed in Figure 13 are not always the same we encountered in the eigenvector centrality (Figure 12). Canada, Italy, the Netherlands, Spain, Russia/USSR are among the nations with the highest values for both eigenvector and joint centrality, but Sweden and Portugal leave their place to Finland and the United States when we shift from the eigenvector centrality to the joint centrality. This change can be accounted for by considering the difference existing between the two measures. While the former takes every year as independent, the second offers a temporal overview. In this temporal overview, nations like Finland and the United States replace Sweden and Portugal because, while having roughly the same total number of members, the members belonging to Finland and the United States were more uniformly distributed during the entire history of the commission (for instance, also after WWII and not just during the interwar years).

## Conditional centrality

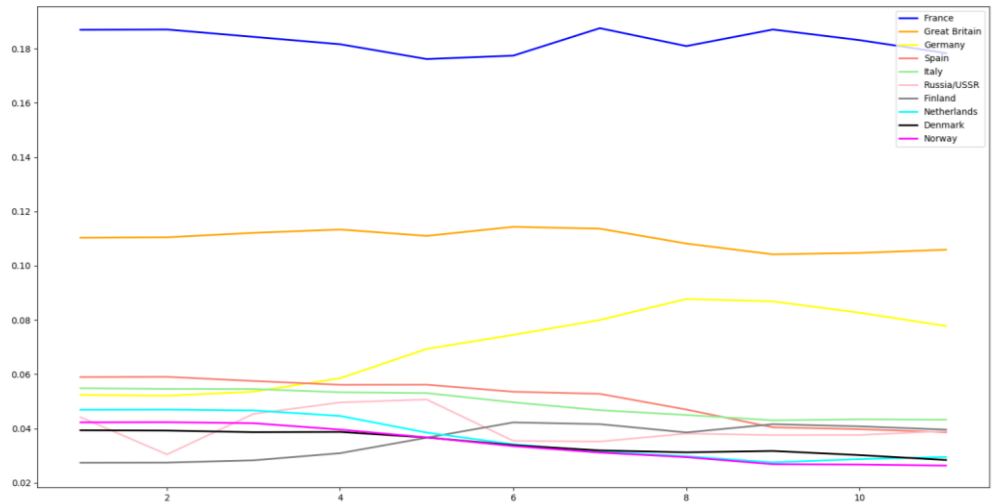

**Figure 14.** Plot of the conditional centrality for nations.

The conditional centrality values substantially confirm what already observed above. Even after normalization of the joint centrality, the resulting values for France, Great Britain, and Germany are the highest. The normalization, however, smooths out the values for the three main nations. For France and Great Britain, this means that the respective plots are centered around a single value (0.18 and 0.11 respectively) with only minimal oscillations, while in the case of Germany, it becomes more evident the troubled history of this country and its scientific establishment during the first half of the twentieth century. Germany's conditional centrality, in fact, raises significantly after 1926, when the post-WWI ban on German scientists was lifted, but again diminishes for the last two meetings of the commission.

The plots of all the other nations represented in Figure 14 are very close one to the other. Again, we have a few differences regarding the nations represented. Compared to Figure 13, here United States and Canada are replaced by Denmark and Norway. As the values for these nations are all very close (they are all grouped in the interval 0.03 to 0.06) we gather that the normalization used to compute the conditional centrality alters slightly the values of the joint centralities modifying marginally the nations' order.

Overall, the study of the national groups done through the centrality measures is limited in scope and interest by the structure of our data. France, Great Britain, and Germany had a larger portion of the overall membership of the commission compared to all the other participating nations and the centrality measures for our network can only confirm this leading position.[27]

## 5        International collaboration and knowledge exchange

The analytical tools of network theory have allowed us to visualize and examine in detail the evolution of the IMO Commission for Agricultural Meteorology, since its appointment in 1913 until its last meeting in 1947. During more than three decades this organization, which came to count more than one hundred and thirty members working in almost forty nations in Europe, Africa, Asia, Australia, and the Americas, promoted knowledge exchange in agricultural meteorology. It did so through its correspondence network which involved all members and was centered around the president and secretary/vice-president of the commission, and through its meetings, which were held regularly but could only be attended by a few members. The commission was a forum for discussing national experiences in agricultural meteorology and for organizing international observational and experimental schemes. It gave members the opportunity to borrow ideas from each other, share their practices, and pursue a standardization of the instruments and methods used in agricultural meteorology. The success of the work done by the commission relied on the collaboration between members, because the IMO was an organization without any diplomatic status and the members were in charge of implementing the IMO resolutions in their respective countries.

For these reasons, we built the network of the IMO technical commission on agriculture taking into account two main elements, i.e. the role of each member in the commission and the nation they worked for. Our analysis has focused on the members that more contributed to the information exchange, that is the presidents and secretaries/vice-presidents, and the long-term members. Among these members we singled out those who held key roles and their long-term influence on the work of the commission. The Swedish hydrographer and meteorologist Axel Wallén emerged as a key presidential figure during the interwar years, but the centrality measures also brought to our attention the value that less prominent individuals, such as the French agronomist Pierre Rey, had in the information exchange of the commission. While interpreting the centralities of individual members, we had to take into account the problems of back causation and the non-null values for members, even after they left the commission. Both issues are intrinsic to the mathematical formalism of network theory and require caution while interpreting the results. Similarly, we noticed that the strong discontinuities existing in our network led to peculiar results and again required care in our interpretation of the centrality measures.

---

[27] Of the 132 total members of the commission, 41 (i.e. 31%) belonged to France, Great Britain, and Germany. The remaining 69% was divided between 34 different nations.

Moving from the members' analysis to the analysis of national groups we had to define new quantifiers, based on the individual members' centrality measures. The nodes in our network, in fact, are not the nations, but the individual members and we could only examine national groups as an aggregate of their members' values. The results of our analysis related to nations offered limited insights due to the overpowering role that three nations – France, Great Britain, and Germany – had compared to all the other nations represented in the commission. The centrality measures we used could only confirm the key role played by these countries and, in the case of Germany, also helped to follow changes in its membership as a consequence of the two world wars. Eigenvector centralities for nations could also point out how the election of a president or a secretary belonging to one of the less represented national groups boosted the national relevance. Overall, however, in a case study in which three nations accounted alone for one-third of the entire commission's membership and the other 34 nations for the remaining seventy per cent, there is little space for a more fine-grained analysis.

As remarked in the introduction to this paper, primary and secondary literature available on the IMO and its technical commission on agriculture is scarce. Yet, using only the membership data gathered from the member lists, we could describe the evolution of this scientific community and its knowledge circulation practices over three decades. We consider this a success of network theory applied to historical sources and believe that this quantitative approach can be especially valuable when sources are scarce, as in our case, or, on the contrary, when sources are too abundant to allow for the traditional 'close reading' historians are accustomed to. As it always happens in research, there is, however, a trade-off to take into account. The network model we built and the interpretation we gave for the centrality measures is grounded in our historical understanding of the commission's work, as it emerges from the sources available, but it must also take into account the issues that the mathematical formalism of network theory generates (for instance, back causation), and come to terms with the potentialities, but also the limits of the analytical tools. In our case, key to finding this balance was the collaborative approach, as this paper has been written together by an historian and a physicist. The former has done the data collection and data cleaning work and has actively participated in the interpretation of the results, while the latter has contributed his technical knowledge of temporal networks, written the code for the data analysis, and completed the computations on which the analysis of the historical case study is based. Both the authors learned a great deal in the process and consider the collaboration profitable.

Returning to the case study, we can only conclude pointing out that scientific communities and institutions are indeed suitable case studies to be approached through network analysis, temporal or not depending upon the circumstances. The main stumbling block, as always in historical research, is the intrinsic difficulty of data collection and data cleaning. On the other hand, when this stumbling block is removed, as in our case due to availability of member lists, network analysis can become a precious asset in the systematic analysis of scientific organizations and their knowledge circulation practices.

## A    Appendix: Mathematical theory of temporal networks

From a mathematical point of view, networks can be described in terms of an adjacency matrix $A$, whose elements $A_{ij}$ are given by:

$A_{ij} = 1$ if nodes labelled $i$ and $j$ share a connection (edge or link)

$A_{ij} = 0$ if nodes labelled $i$ and $j$ are not connected.

If the network is weighted, that is some connections have a weight greater than 1, say $w$, then the above expression is replaced by:

$A_{ij} = w$ if nodes labelled $i$ and $j$ share a connection (edge or link)

$A_{ij} = 0$ if nodes labelled $i$ and $j$ are not connected.

A weight can be the number of times two nodes interact along the history of the network or any other convenient measure.

The adjacency matrix is a square matrix of dimension N×N. In this matrix all the elements along the diagonal ($A_{ii}$) are equal to zero, since nodes are not connected to themselves.

By mapping a network onto an adjacency matrix, one can use well-known results from linear algebra to extract relevant information about the matrix and therefore about the network itself. Among these measures the most intuitive one is the so-called degree centrality or simply degree of a node, defined as:

$$degree\ of\ node\ i = \sum_{j \neq i} A_{ij}$$

That is, the degree of node $i$ ($i$ is the row index) is the sum over the column index $j$ of all matrix elements of the form $A_{ij}$ where $j \neq i$. The degree of a node measures the number of incoming/outgoing links. One might think that the more links a node has, the more relevant it is, however, in network theory a node with many connections is not necessarily an important one, since it might be connected to irrelevant nodes. To better assess the relevance of a node one can define the so-called eigenvector centrality, which is given by the entries of the eigenvector associated with the largest eigenvalue λ of matrix $A$. Mathematically speaking it is given by the elements of the vector matrix $v$ of dimension N×1 which satisfies the following eigenvector equation:

$$Av = \lambda v$$

The eigenvector centrality of a given node is related to its degree, but it also takes into account the 'quality' of the nodes to which that particular node is connected.

Centralities are obtained by counting the elements of a network (e.g. links to/from a node; shortest paths passing through a node, etc.). Due to their nature, they can only be positive numbers. This requirement might become problematic when calculating the eigenvector centrality, because the entries of a vector $v$ might be negative. In this case one has to recur to the Perron-Frobenius Theorem, which asserts that if the matrix is square and symmetric, i.e. $A_{ij} = A_{ji}$ then the eigenvector centralities are all positive. This condition is automatically satisfied by all undirected networks, as the one discussed in the present paper. When it comes to directed networks, some care is required. The Perron-Frobenius Theorem can still be used but one has to satisfy some more stringent conditions to guarantee the positivity of the eigenvector centralities. As in our case the Perron-Frobenius Theorem applies, we will not delve deeper into these details.

## Temporal Networks

All network measures can be defined irrespective of whether the network changes with time or not. The most direct way to analyze a network that changes over time is to take the different realizations of the network, i.e. $A_1$, $A_2$, $A_3$, ..., $A_M$ for different time frames $t$ = 1, 2, 3, ... M, and add all these matrices to form what is called an aggregate matrix $A$:

$$A = A_1 + A_2 + A_3 + A_1 + \cdots + A_M$$

However, some network features may strongly depend on the time factor, and the direct sum above means basically that some nodes that left the network at some point will still be represented as if they were still part of it. In other words, the temporality is lost. Therefore, it is important to find ways to represent a series of time-dependent adjacency matrices in a way to keep the temporality features but still be able to use the regular matrix measures known for networks represented by just one matrix. There are several techniques to accomplish this and the choice of which technique to use is contingent on the question being asked. Of all these flattening techniques, as they are called, in this paper we opted to use one developed by Taylor et al. which is based on defining a superadjacency matrix $A_T$ as follows:[28]

$$A_T = \begin{pmatrix} \boldsymbol{A_1} & \boldsymbol{\alpha_1} & \boldsymbol{0} & \cdots \\ \boldsymbol{\alpha_1} & \boldsymbol{A_2} & \boldsymbol{\alpha_1} & \cdots \\ \boldsymbol{0} & \boldsymbol{\alpha_1} & \boldsymbol{A_3} & \cdots \\ \vdots & \vdots & \vdots & \ddots \end{pmatrix}$$

For instance, if $A$ is a 3×3 matrix (three nodes) and there are two time steps:

$$\begin{pmatrix} A_{11}^1 & A_{12}^1 & A_{13}^1 & \boldsymbol{\alpha} & 0 & 0 \\ A_{21}^1 & A_{22}^1 & A_{23}^1 & 0 & \boldsymbol{\alpha} & 0 \\ A_{31}^1 & A_{32}^1 & A_{33}^1 & 0 & 0 & \boldsymbol{\alpha} \\ \boldsymbol{\alpha} & 0 & 0 & A_{11}^2 & A_{12}^2 & A_{13}^2 \\ 0 & \boldsymbol{\alpha} & 0 & A_{21}^2 & A_{22}^2 & A_{23}^2 \\ 0 & 0 & \boldsymbol{\alpha} & A_{31}^2 & A_{32}^2 & A_{33}^2 \end{pmatrix}$$

In this example we can see that, for instance, the element $(A_T)_{14}$ of the matrix $A_T$ connects the node represented by $A^1{}_{11}$ during the first realization of the network with its own image $A^2{}_{11}$ at the following time step. The coupling $\alpha$ represents how strongly different time layers are connected. A value of $\alpha$ close to zero means that the submatrices are not connected at all and the matrix $A_T$ breaks down into two independent matrices.

If one has M different time steps, each separate matrix $A_k$ represents the adjacency at that particular time layer (in our present study M=11, each time step representing one year for which a member list is available). The quantities are defined as usual for the new superadjacency matrix $A_T$. However, one has to proceed with care. If one calculates the eigenvector centrality of this new matrix of dimension (NM)×(NM), there will be N×M different values where the first N entries correspond to the eigenvector centralities of the nodes at the first realization of the network, the second N entries at the second realization and so on. Note that these values are different from the eigenvectors of each realization calculated separately, that is for each $A_k$ (k = 1, 2, 3, ... M).

To avoid any confusion, the eigenvector centralities obtained from $A_T$ are called joint centralities to indicate that each value is related to a specific node $i$ at a particular time $t$. We call this quantity $W_{it}$. It is given by the entries of the eigenvector $\mathbf{v}_T$ of the matrix $A_T$ as follows:

$$A_T v_T = \lambda_T v_T \rightarrow W_{it} = (v_T)_i(t)$$

---

[28] Taylor et al., "Eigenvector-Based Centrality Measures for Temporal Networks". With $\alpha_1$ we indicate the identity matrix multiplied by the coupling constant $\alpha$.

For each node $i$, the centrality value corresponds to all those entries in position $i \times t$ of the vector $v_T$. So for example the joint centrality of node $i$=5 at time step $t$=6 is to be found as the 30th element of the $\mathbf{v}_T$ vector.

For a given time we can measure also the relative importance of a node within that particular time layer by dividing its joint centrality by the sum of the centralities of all the other nodes. If for a given $t$ we take the sum:

$$y_t = \sum_i W_{it}$$

then we can define the conditional centrality $Z_{it}$ as:

$$Z_{it} = \frac{W_{it}}{\Sigma_i W_{it}} = \frac{W_{it}}{y_t}$$

Once the matrix $\mathbf{A}_T$ is built, one can use any computational package to manipulate it. Multiple programming languages such as C++ and Fortran are suitable for the task, but it is becoming increasingly popular to use Python routines for handling networks. The network measures used in this paper—eigenvector centralities, joint centralities, and conditional centralities—were all obtained using code written in Python.[29]

---

[29] The Python code has been developed by Sílvio Dahmen (the guidance of Giovanna Lazzari Miotto with Python is gratefully acknowledged).